# Metastable defects in III-Nitrides

A.E.Belyaev

*V.Lashkaryov Institute of Semiconductor Physics, NASU, Kiev, Ukraine*

*Abstract.*

Metastable states in III-nitrides (AlN, GaN, InN and their alloys) are an important topic across materials science, device physics, and epitaxial growth because these compounds exhibit multiple competing crystal structures, defect configurations, and phase behaviors that strongly affect optical, electrical, and mechanical properties. The paper is a concise, structured review-style summary covering the main concepts, mechanisms, experimental signatures, theoretical approaches, and device implications. Actions that should be proposed to avoid effect of metastable defects in III-Nitride based devices are discussed. Avoiding metastable defect effects in III-nitride devices requires preventing their formation (through growth and doping control), stabilizing their electronic configuration (by annealing or optical activation), and designing device architectures that minimize field- or temperature-induced transitions. These combined actions ensure consistent carrier density, higher mobility, and superior optical and electrical reliability in GaN-, AlGaN-, and InGaN-based devices.

## Intrinsic and extrinsic metastable centers.

Metastable defects in III-Nitrides (GaN, AlGaN and GaN/AlGaN heterostructures) are defect states that can exist in more than one stable configuration, often depending on factors like temperature, illumination, or charge state. These defects significantly influence the optical and electronic behavior of III-Nitrides, which is crucial for electronic and optoelectronic devices such as high electron mobility transistors (HEMTs), light emitting diodes (LEDs) and laser diodes (LDs).[1] Research has identified both intrinsic (native) and extrinsic (impurity-related) metastable centers.[2] First, let's consider gallium nitride's possibilities for forming metastable states. Native defects such as gallium and nitrogen vacancies are among the most studied.[3,4] Nitrogen vacancies ($V_N$) can act as shallow donors, while Ga vacancies ($V_{Ga}$) act as deep acceptors. Both can form metastable complexes under different charge states. Recent DFT and hybrid-functional calculations show that $V_N$ can exhibit negative-U behavior, meaning one charge state (e.g., 2+) is unstable, and transitions skip it — a hallmark of metastability.[5] Experimentally, metastable defects associated with Ga vacancies were found to produce blue and yellow luminescence (YL) bands, which switch intensity depending on temperature and excitation. These centers may trap holes and show luminescence fatigue — they lose intensity under continuous illumination but recover at room temperature.

Expanding the functionality of III-nitride-based devices requires the use of intentional doping. However, impurity-related metastable defects may form in this case. For example, Eu-doped Gallium nitride (GaN) is a promising candidate for GaN-based red light-emitting diodes, which are needed for future micro-display technologies. Introducing a superlattice structure comprised of alternating undoped and Eu-doped GaN layers has been observed to lead to an order-of-magnitude increase in output power.[6] Europium ($Eu^{3+}$) introduces metastable defect complexes, specifically ($Eu_{Ga}$–$V_N$). Ab initio calculations reveal two stable configurations — axial and basal — each with distinct energies and electronic levels. These states can switch depending on the charge state of the defect, producing charge-state controlled metastability.[7,8]

Magnesium (Mg) is the most commonly used p-type dopant for GaN.[9,10] Mg-related metastable behavior is observed in near-bandgap UV luminescence, where the luminescence spectrum evolves under electron irradiation. The process is reversible with heat, attributed to stacking-fault–related defect activation — suggesting metastability stems from irradiation-induced defect rearrangement.[11] It should be noted how the above-mentioned defects manifest themselves. First, several broad photoluminescence (PL) bands (e.g., yellow, blue, green) in GaN are linked to metastable defect transitions. Second, metastability leads to temperature- and light-dependent emission switching, persistent photoconductivity, and luminescence fatigue effects. As a result, understanding these centers is crucial for controlling nonradiative recombination and improving GaN-based devices' efficiency. In essence, metastable defects in GaN arise from both intrinsic vacancies and impurity-defect complexes (like Eu- or Mg-related). They exhibit charge-state or configuration-dependent bistability that affects optical emissions and carrier dynamics — making their characterization vital for high-performance photonic and electronic applications.

Recently, a lot of attention has been paid to impurities, which, along with hydrogen-like states, can form states with negative correlation energy, the so-called DX states. DX centres have first been identified in AlGaAs alloys and the very detailed studies of donors in these alloys are the most relevant benchmarks for the characterisation and the theoretical models used to understand DX behaviour in other materials.[12]

In the AlGaN alloy system, three initial investigations of the DX behaviour of donors have been published.[13-15]

Silicon (Si) on Ga sites ( $Si_{Ga}$) is most studied DX-forming dopant in GaN. In pure GaN at ambient pressure, silicon typically acts as a shallow donor (responsible for n-type conductivity). Under high pressure, high Al content in $Al_xGa_{1-x}N$,or compensating conditions, $Si_{Ga}$ can undergo a DX transition, producing a deep, metastable configuration with strong localization of one electron and lattice distortion. Negative-U behavior observed in Al-rich alloys — the DX center becomes more favorable as Al content increases beyond $x \approx 0.4$–$0.5$ in AlGaN.[16] Si is an established shallow n-type dopant in GaN and $Al_xGa_{1-x}N$ ($x<0.8$). However, an increased ionization energy is observed for Si in $Al_xGa_{1-x}N$ (for $x>0.8$) that is commonly attributed to the formation of a DX center, consequently limiting the achievable conductivity in high Al content AlGaN.[17-20]

Germanium (Ge) acts as a shallow donor in GaN and is sometimes used as an alternative to Si. Under high pressure or in Al-rich environments, Ge-induced DX formation has been predicted theoretically, but experimentally, Ge remains largely stable and shallow in GaN. The DX configuration becomes more relevant in $Al_xGa_{1-x}N$ alloys for $x > 0.5$, similar to Si.The behavior of the Ge dopant in AlGaN was studied as a function of Al-content in the alloy. At Al compositions below 40%, it behaved as a shallow donor with ionization energy below 20meV. At Al compositions above 40%, it emerged as a deep donor, whose ionization energy increased with increasing Al content and reached 150meV for AlGaN with 60% Al. At this point, only the deep donor state was observed.[21]

Oxygen is a common unintentional donor in GaN. First-principles studies predict that $O_N$ can form DX-like centers at high Al concentration in AlGaN alloys.[22,23] In GaN itself, $O_N$ is generally shallow, but in AlGaN with Al > ~30–40%, it may behave as a deep DX-like donor due to conduction-band lowering and local relaxation effects.

Heavy donors like Te and Se have been suggested (by first-principles studies) to induce deeper DX-like configurations even in GaN, but experimental verification is scarce.[24,25] These dopants are rarely used due to incorporation challenges and amphoteric behavior.

As a result, in pure GaN, classic DX centers are rare — most donors (Si, O, Ge) remain shallow.

However, as the local environment changes (pressure, Al incorporation, strain, or high electric field), these donors can transform into metastable, deep DX-like states, strongly affecting carrier compensation and device reliability. In Table I DX center formation trends are summarized.

Table I.

| ***Dopant*** | ***Site*** | ***DX Formation in GaN*** | ***DX Formation in AlGaN*** | ***Notes*** |
|---|---|---|---|---|
| Si | Ga | Weak / under pressure | Strong (x > 0.4) | Most studied case |
| Ge | Ga | Minimal | Appears for high Al fractions | Alternative to Si |
| O | N | Weak / metastable | Strong (x > 0.3) | Unintentional donor |
| Te, Se | N | Theoretical only | Possible | Not widely verified |

In aluminum nitride (AlN), DX centers are theoretically and experimentally predicted to be even more stable and deeper than in GaN because of AlN's wider bandgap (~6.2 eV), stronger ionicity, and larger lattice relaxation energy. The presence and behavior of DX centers in AlN are therefore crucial for understanding the electrical compensation and n-type doping limits in Al-rich alloys ($Al_xGa_{1-x}N$). A DX center forms when a donor impurity undergoes a large lattice relaxation after capturing an electron, moving to an off-center position and creating a deep-level configuration. Because AlN is more ionic and has a deeper conduction band minimum (CBM) than GaN, donors that are shallow in GaN often become deep or unstable (DX-like) in AlN.[26-30] In Table II likely DX-forming dopants are presented.

Table II.

| ***Dopant*** | ***Site*** | ***Behavior in AlN*** | ***Notes*** |
|---|---|---|---|
| Silicon (Si) | Al site | Forms deep DX-like state | Shallow in GaN, but deepens in $Al_xGa_{1-x}N$ for $x > 0.4 – 0.5$; in pure AlN, typically deep (> 0.4 eV). |
| Germanium (Ge) | Al site | Deep donor / possible DX | Similar to Si, predicted to form off-center configuration; limited activation experimentally. |
| Oxygen (O) | N site | Strong DX former | $O_N$ becomes deep (≈1–2 eV below CBM); explains self-compensation and limited n-type doping. |
| Sulfur (S) | N site | Deep-level DX | Predicted via first-principles to form deep donor levels > 1 eV below CBM. |
| Tellurium, Selenium (Te, Se) | N site | Deep and unstable | Theoretical DX character; experimental realization very poor. |

The DX configuration corresponds to a large displacement (~0.4 Å) of the donor atom toward a neighboring nitrogen atom, creating a broken bond and localized electron. Negative-U behavior is common: the singly charged state ($d^0$) is unstable, and the defect transitions directly between $d^+$ and $D^-$ due to lattice relaxation. Calculated capture barrier energies range from 0.2 – 1.0 eV, contributing to strong metastability and persistent photoconductivity effects. First-principles studies[22,23] predict $O_N$ and $Si_{Al}$ donors to form deep DX levels in AlN with transition levels 1–2 eV below the conduction band. The manifestation of DX centers is observed in optical and electrical measurements (AlN doped with Si or O shows very low free-electron concentrations despite high impurity levels, consistent with self-compensation by DX centers and electron trapping) as well as in photo- and thermally stimulated conductivity (at elevated temperatures or under illumination, partial ionization of DX centers increases conductivity, but equilibrium returns upon cooling or in the dark). The presence of DX centers significantly affects on device physics. DX formation is the main reason AlN is extremely difficult to dope n-type — most donors become deep and inactive. The crossover from shallow to DX behavior as Al content increases defines the “DX transition composition” (typically 40–60% Al). Deep DX states can act as non-radiative recombination centers and trap states, affecting UV LED efficiency and reliability.
In summary: DX centers in AlN are deep, strongly bound donor configurations—primarily involving Si and O impurities—that dominate compensation processes and limit n-type conductivity. Their formation energy decreases with Al content, making AlN and high-Al AlGaN alloys intrinsically prone to DX-related metastability and deep donor behavior.

**Effect of DX centers on properties of 2DEG in AlGaN/GaN heterostructures.**
In AlGaN/GaN heterostructures, DX centers—mainly those associated with donor impurities such as Si or O in the AlGaN barrier—have a substantial impact on the two-dimensional electron gas (2DEG) that forms at the interface. These centers affect both the formation and the stability of the 2DEG by altering charge supply, local electrostatics, and trap dynamics. It is well established that 2DEG arises from:

- The polarization discontinuity (spontaneous and piezoelectric polarization) between AlGaN and GaN, creating a built-in electric field.
- Donor impurities (intentionally added or residual, e.g., Si, O) in the AlGaN barrier, which act as charge reservoirs supplying electrons to the interface.

First of all, one should consider the mechanisms by which DX centers affect the 2DEG. Among them are:

***a. Charge supply modulation***

- Some dopants (especially Si and O) in the AlGaN barrier can exist in two states:
  - Shallow donor ($d^0$): provides electrons to the 2DEG.
  - Deep $DX^-$ state: traps electrons, unavailable for conduction.
- When a large fraction of donors convert into the DX state, the effective donor density drops, lowering the sheet carrier density of the 2DEG.
- At low temperatures or under bias stress, these DX centers can remain negatively charged, pinning or depleting the 2DEG.

***b. Metastability and dynamic trapping***

- The conversion between shallow and deep states is thermally or optically activated:
  - Cooling can freeze donors in the $DX^-$ configuration, reducing 2DEG density.
  - Illumination or heating releases trapped electrons, increasing it again.
- This leads to persistent photoconductivity (PPC) or bias-stress hysteresis, revealing the metastable nature of DX centers.

***c. Electric-field interaction***

- The large polarization field in the AlGaN barrier (~1–2 MV/cm) favors DX formation because it enhances local lattice relaxation around donors.
- During high-bias operation, field-induced electron emission from DX centers can cause threshold shifts or transient current increases.

***d. Scattering and mobility degradation***

- Ionized or neutral DX centers act as Coulombic scattering sites, reducing mobility, especially when their distribution is near the 2DEG channel.
- They also contribute to current collapse and dispersion in high-frequency devices (e.g., HEMTs), through trap-assisted charge modulation.

From the currently available literature sources, it is possible to summarize the effects observed in AlGaN/GaN heterostructures (Table III).

Table III.

| ***Property*** | ***Typical observation*** | ***DX-related interpretation*** |
|---|---|---|
| 2DEG density vs temperature | Decreases at low T, recovers after illumination | Electrons trapped in $DX^-$ states |
| Persistent photoconductivity | Long-lived conductivity after UV or blue light | Photoionization of DX centers |
| Current collapse / gate lag | Transient current reduction after bias stress | Electron trapping in or near barrier DX centers |
| Hysteresis in C–V or I–V curves | Especially under slow sweep rates | Thermal recovery of metastable donors |
| Drain-induced barrier lowering artifacts | Time-dependent threshold drift | Field-assisted DX ionization |

Based on this, we can make several recommendations regarding reducing the impact of metastable states, namely:

- Use of undoped or delta-doped barriers: reduces DX-related charge trapping.

-Lower Al content: below ~40%, Si donors stay mostly shallow, minimizing DX formation.

-Thermal or optical activation: can intentionally reset or “de-trap” electrons from DX states.
-Alternative dopants (e.g., Ge): shown to produce more stable, shallow donor behavior in mid-Al-content AlGaN.
In summary: DX centers in the AlGaN barrier act as metastable charge traps that dynamically control the electron transfer into the GaN channel. Their presence can lower carrier density, cause persistent photoconductivity or transient effects, and degrade mobility and reliability in AlGaN/GaN HEMT structures—especially in high-Al-content barriers where DX behavior is more pronounced.

**The role of external factors.**

It is very important to consider the role of external factors, such as ultrasonic vibrations or microwave (including terahertz) radiation, on the effects associated with DX centers.
DX centers are strongly electron-phonon coupled systems; ultrasound effectively introduces a *coherent phonon population*. If the acoustic energy is comparable to a fraction of the DX barrier height (tens to hundreds of meV), it can significantly alter the equilibrium between the deep and shallow states, especially at moderate temperatures (~100–300 K).
**Ultrasonic vibrations** can significantly influence DX centers in GaN and related III–V semiconductors by providing mechanical energy that couples to the local lattice strain and electronic configuration of the defect. The underlying mechanism is tied to how DX centers involve a strong electron–lattice relaxation—the donor atom moves off-center and distorts nearby bonds when it captures electrons. Ultrasonic waves introduce a periodic stress field that perturbs these local distortions.
Here’s how the effect works in more detail:

**Strain coupling and potential-energy modulation.**

A DX center has two (or more) stable configurations:

- Shallow donor ($d^0$) - donor ion in its substitutional position, hydrogen-like shallow level.
- Deep $DX^-$ state - donor atom displaced off-center, lattice distorted, deep localized level.

The two are separated by an energy barrier on the configuration coordinate diagram. Ultrasonic strain modulates the local potential energy landscape. Acoustic pressure periodically alters the lattice spacing, changing bond angles and distances near the dopant (Si, O, etc.). This modulation can raise or lower the barrier between $d^0$ and $DX^-$ states and even cause thermally assisted transitions between them. As a result, populations of shallow and deep states change dynamically — modifying carrier concentration and conductivity.

**Acoustically stimulated ionization and annealing.**

At suitable frequencies (typically MHz range) and power levels the piezoelectric field generated by ultrasound in GaN adds an oscillating electric field component that can assist ionization of electrons from $DX^-$ back to the conduction band. This can temporarily increase free-electron concentration and reduce compensation. Over prolonged exposure, strong ultrasonic fields can assist defect migration or annealing, promoting DX-to-shallow donor conversion or even structural healing of the local distortion.

**Observable effects**

Several measurable consequences have been reported or predicted:

- Changes in resistivity: Ultrasonic excitation can decrease resistivity in n-type GaN by releasing electrons trapped in DX centers.
- Altered luminescence intensity: PL or CL (cathodoluminescence) can show reversible intensity changes as the occupation of deep centers fluctuates.
- Acoustic attenuation and velocity changes: The interaction between DX centers and strain fields leads to characteristic temperature- and frequency-dependent attenuation peaks (acoustic relaxation).
- Persistent photoconductivity modulation: Ultrasonic waves can accelerate recovery from persistent photoconductivity by providing an alternative relaxation channel for metastable DX defects.

In short, ultrasonic vibrations affect DX centers by modulating strain, electric fields, and defect potential landscapes, leading to reversible changes in charge state, conductivity, and optical activity.

This coupling between acoustics and defect dynamics is exploited in acoustic annealing experiments and in studies of lattice-defect energetics in GaN and AlGaN materials.

**Effect of high frequency (including terahertz) irradiation on properties of GaN with DX centers.**

High-frequency electromagnetic irradiation—including microwave, sub-terahertz, and terahertz (THz) ranges—can have marked effects on GaN containing DX centers, because these metastable donor defects couple strongly to both electric fields and lattice vibrations. The response depends on the field amplitude, photon energy, and the defect's configuration barrier (typically tens to hundreds of meV). Let's look at possible mechanisms of interaction:

***a. Electric-field coupling (microwave to THz):***

- DX centers involve a polar lattice distortion—the donor atom (e.g., $Si_{Ga}$ or $O_N$) shifts off-center in its lattice site.
- Alternating electric fields from high-frequency radiation periodically polarize the lattice, modulating this off-center displacement.
- Strong fields can induce charge-state transitions ($DX^- \leftrightarrow d^0$) by field-assisted tunneling or ionization of the trapped electron into the conduction band.
- The ionization threshold corresponds to photon energies of about tens of meV (≈ THz range), so THz radiation can resonantly excite bound electrons in shallow or metastable states.

***b. Electron-heating mechanism:***

- Under sub-bandgap irradiation (e.g., intense THz pulses), conduction electrons gain kinetic energy ("hot electrons") without generating new carriers by absorption.
- Hot electrons can thermally detrap from DX centers, raising the free electron concentration and transient conductivity.
- Relaxation after the pulse can reveal recovery kinetics of the DX state.

***c. Phonon resonance and lattice coupling:***

- Terahertz frequencies overlap with optical phonon modes.
- Intense THz fields can drive nonlinear phonon oscillations, locally altering the lattice potential around DX centers and transiently flattening or reshaping the energy barrier between shallow and deep configurations.

**What are observable effects and phenomena in this case:**

- Photoionization of DX centers: THz photons with $\gtrsim 30–100\,\text{meV}$ energy can ionize deep donor electrons, similar to mid-IR radiation but without creating band-to-band transitions.
- Conductivity changes: Measured as transient increases in carrier density or THz transmission (time-domain THz spectroscopy).
- Metastability control: Repetitive THz exposure can depopulate $DX^-$ centers, shifting material toward its shallow-donor configuration—effectively *"photo-annealing"* the metastable state.
- Luminescence modulation: Band-edge or deep-level PL intensity can change due to altered defect charge states.
- THz absorption signatures: DX-related transitions produce characteristic sub-bandgap absorption or relaxation peaks, useful for defect spectroscopy.

In Table IV we *c*ompare manifestation of different effects across frequency regimes.

Table IV.

| *Frequency Range* | *Dominant Interaction* | *Typical Outcome* |
| --- | --- | --- |
| MHz–GHz (microwave) | Field-assisted tunneling, local heating | Minor conductivity modulation |
| 0.1–3 THz | Photon-assisted ionization, electron heating, phonon coupling | $DX^- \rightarrow d^0$ conversion, photoconductivity rise |
| >10 THz (far-IR) | Resonance with LO-phonons, strong lattice coupling | Structural relaxation, possible persistent state change |

Several practical applications of the effects discussed above can be suggested:

- Reversible control: Non-destructive tuning of donor occupation and compensation level via THz pulses.
- Characterization tool: THz time-domain spectroscopy can probe DX-related trap energies and capture/emission times.
- Potential degradation or annealing: Continuous high-intensity THz or microwave exposure could either stabilize the shallow configuration (reducing compensation) or, at excessive power, damage the lattice through overheating.

Summarizing this paragraph, the following conclusion can be drawn:
Terahertz and other high-frequency fields can dynamically control the electronic and structural state of DX centers in GaN by driving charge emission, lattice relaxation, and local polarization switching. These processes alter conductivity, optical emission, and metastability, offering both diagnostic and defect-engineering routes for wide-bandgap semiconductors.

**Effect of DX centers on properties of plasmonic structures based on AlGaN/GaN heterostructures.**

DX centers in AlGaN/GaN heterostructures strongly influence the optical and electronic behavior of plasmonic structures, especially those exploiting the 2DEG (two-dimensional electron gas) at the interface as the plasmonic medium. Because plasmonic resonances in these systems depend sensitively on carrier density, scattering rates, and local permittivity, DX centers—acting as metastable donors and traps—affect both the strength and dynamics of plasmonic response.
First, we should consider the effects of the DX centers on the plasmonic response. DX centers in AlGaN (mainly $Si_{Ga}$ or $O_N$) control how many electrons are donated to the 2DEG. When donors convert into $DX^-$ states, fewer electrons populate the 2DEG and plasmon frequency decreases and intensity weakens. Under optical or high-field excitation, ionization of $DX^- \rightarrow d^0$ releases electrons and plasmon frequency blue-shifts and amplitude increases. This is the basis for photo- or bias-tunable plasmonic effects in GaN-based heterostructures.
Nonlinear and time-dependent behavior must also be taken into account. DX centers are metastable; transitions between charge states are slow (μsec–sec scale). This produces plasmonic hysteresis and memory effects—plasmon resonances depend on illumination history or bias cycling. Illumination can produce persistent photoplasmonic conductivity, analogous to persistent photoconductivity, where plasmon resonances persist long after light is turned off.
Finally, two more important factors should be considered when analyzing experimental results. First, **mobility degradation**. Charged DX centers act as strong Coulomb scattering centers near the interface. They increase electron scattering rate ($\tau$), broadening and damping plasmon peaks. As Al content rises ($x > 0.4$), DX centers become deeper and more abundant, reducing plasmonic quality factor $Q = \omega_p / \tau$ . Second, l**ocal permittivity and field uniformity.** Spatially inhomogeneous charge trapping at DX sites causes fluctuations in the local dielectric environment. This leads to plasmon resonance broadening and mode localization, particularly in nanostructured metasurfaces or gratings fabricated on AlGaN/GaN. Table V summarizes observable consequences in this case.

Table V.

| ***Property*** | ***DX-related influence*** | ***Experimental manifestation*** |
|---|---|---|
| THz plasmon resonance frequency | Red-shift when $DX^-$ dominant; blue-shift after photoionization | Reversible tuning under UV/THz illumination |
| Linewidth / Q-factor | Increased due to carrier scattering | Broadened spectral response |
| Intensity (plasmon coupling strength) | Reduced by carrier depletion | Lower transmission/reflectance modulation |
| Time response / hysteresis | Slow recovery (~sec–min) due to trapped carriers | Memory effect in THz or IR modulation |
| Temperature dependence | Recovery of $n_{2DEG}$ with heating ($DX^- \rightarrow d^0$ re-ionization) | Thermally activated resonance shift |

In summary, DX centers in the AlGaN barrier of plasmonic AlGaN/GaN heterostructures act as metastable charge traps that dynamically control the 2DEG density, scattering rate, and local dielectric environment. They cause frequency shifts, intensity modulation, and damping of plasmon resonances, while also enabling persistent and reversible control via optical or electrical stimuli. This dual role makes DX centers both a source of loss and a potential mechanism for active tuning or memory in GaN-based plasmonic and terahertz devices.

**What actions should be proposed to avoid effect of metastable defects in III-Nitride based devices.**

Suppressing or mitigating the effects of metastable defects (like DX centers, vacancy complexes, or impurity-related traps) in III-nitride devices (GaN, AlGaN, InGaN, AlN) requires a combination of material engineering, process optimization, and device design strategies. The goal is to minimize defect formation, prevent their activation, and reduce their impact on charge transport, optical recombination, and long-term stability. Five steps are proposed to implement this task:

**Step 1. Materials growth strategies**

***a. Optimize growth conditions***

- Maintain near-stoichiometric conditions and low background impurity levels during MOCVD or MBE growth to minimize formation of nitrogen vacancies and unwanted dopant complexes.
- Avoid excessively high Al content and high pressure, which favor lattice relaxation leading to DX states in n-type dopants (especially Si and O).

***b. Use alternative doping strategies***

- Ge instead of Si for n-type doping: Ge donors stay shallower and show less DX tendency in AlGaN with moderate Al fractions ($x < 0.6$).
- Minimize O and H contamination, which form metastable or amphoteric centers contributing to compensation.
- For p-type doping, control Mg–H complex formation (a metastable pair) through proper activation annealing.

***c. Heterostructure design adjustments***

- Use lower Al composition (or graded Al profiles) in the barrier layers of AlGaN/GaN heterostructures to stay below the DX transition threshold (typically 40–50% Al).
- Introduce spacer layers or undoped buffer regions to spatially separate dopant atoms from active regions (reduces trap scattering and charge transfer variability).

**Step 2. Post-growth treatments**

***a. Thermal annealing***

- Proper post-growth annealing (700–900 °C) helps convert metastable defect configurations back into their shallow donor states and remove hydrogen passivation.
- Multi-step anneals under nitrogen or ammonia can reduce vacancy complex density and stabilize dopant charge states.

***b. UV or optical activation***

- Controlled UV illumination can empty deep traps, restore free-carrier concentration, and “reset” devices before operation.
- Used cautiously, this can pre-stabilize the lattice and minimize drift during device use.

**Step 3. Defect passivation and compensation control**

- Deposit dielectric caps ($SiN_x$, $Al_2O_3$, $HfO_2$) to suppress surface states that interact with internal metastable centers.
- Apply hydrogen control: while H can passivate defects, excess hydrogen introduces metastable Mg–H and donor–H complexes; carefully balance through annealing or plasma treatments.
- Introduce compensating dopants or co-doping schemes to reduce Fermi-level pinning that promotes DX conversion.

**Step 4. Device and operational strategies**

***a. Limit high electric fields and bias stress***

- Avoid prolonged high-field operation or bias stress that drives charge transfer into deep DX states.
- Design with field plates or graded barriers to lower peak fields around gate edges in HEMTs.

***b. Stabilize thermal environment***

- Thermal cycling enhances reconfiguration of metastable centers. Maintain steady operating temperatures or provide self-heating management via efficient thermal paths.

***c. Preconditioning before use***

- Controlled electrical or optical conditioning helps “fill” or “empty” traps in a predictable way, improving reproducibility.

**Step 5. Characterization and feedback control**

- Employ deep-level transient spectroscopy (DLTS), cathodoluminescence, and time-resolved photoconductivity to identify metastable traps and monitor their evolution during processing.
- Use this feedback to fine-tune growth and annealing parameters for each specific device structure.

Table VI summarizes the results of performing the steps listed above.

Table VI.

| ***Approach*** | ***Purpose*** | ***Key Benefit*** |
|---|---|---|
| Low-Al barrier or graded AlGaN | Prevent DX formation | Stable 2DEG density |
| Ge doping (instead of Si) | Avoid DX centers | Shallow donor behavior |
| Post-growth anneal | Relax metastable configurations | Improve carrier activation |
| Controlled hydrogen management | Prevent metastable hydrogenic defects | Long-term stability |
| Field-management design | Reduce charge trapping | Mitigate current collapse |
| Optical pre-stimulation | Reset defective charge states | Stable initial operation |

**General conclusion:**

The paper is a concise, structured review-style summary covering the main concepts, mechanisms, experimental signatures, theoretical approaches, and device implications. Actions that should be proposed to avoid effect of metastable defects in III-Nitride based devices are discussed. Avoiding metastable defect effects in III-nitride devices requires preventing their formation (through growth and doping control), stabilizing their electronic configuration (by annealing or optical activation), and designing device architectures that minimize field- or temperature-induced transitions. These combined actions ensure consistent carrier density, higher mobility, and superior optical and electrical reliability in GaN-, AlGaN-, and InGaN-based devices. It is shown that suppressing or mitigating the effects of metastable defects (like DX centers, vacancy complexes, or impurity-related traps) in III-nitride devices (GaN, AlGaN, InGaN, AlN) requires a combination of material engineering, process optimization, and device design strategies. The goal is to minimize defect formation, prevent their activation, and reduce their impact on charge transport, optical recombination, and long-term stability.

**Acknowledgments**

This study was supported by the National Academy of Sciences of Ukraine (Project No. 5.2/26-П). A.E.B. gratefully acknowledges the long- term program supporting the Ukrainian research teams at the Polish Academy of Sciences, which was carried out in collaboration with the U.S. National Academy of Sciences with the financial support of external partners (Project: LTP NAS

KOCHELAP A7.11.0008; contract # PAN.BFB.S.BWZ.367.022.2023). The authors would also like to thank V.A. Kochelap and V.V. Koroteyev for the valuable collaborative work.

***References.***